\documentclass[sigconf]{acmart}
\AtBeginDocument{%
  }

\setcopyright{acmlicensed}
\copyrightyear{2026}
\acmYear{2026}
\setcopyright{cc}
\setcctype{by-nc-nd}
\acmConference[CHI EA '26]{Extended Abstracts of the 2026 CHI Conference on Human Factors in Computing Systems}{April 13--17, 2026}{Barcelona, Spain}
\acmBooktitle{Extended Abstracts of the 2026 CHI Conference on Human Factors in Computing Systems (CHI EA '26), April 13--17, 2026, Barcelona, Spain}
\acmDOI{10.1145/3772363.3798973}
\acmISBN{979-8-4007-2281-3/2026/04}

\usepackage{booktabs}   
\usepackage{graphicx}   

\begin{document}

\title{IntentWeave: A Progressive Entry Ladder for Multi-Surface Browser Agents in Cloud Portals}

\author{Wanying Mo}
\authornote{The corresponding author.}
\orcid{0009-0009-9426-2426}
\affiliation{%
  \institution{Alibaba Cloud Computing\\Alibaba Group}
  \city{Hangzhou, Zhejiang}
  \country{China}
}
\email{mowanying.mwy@alibaba-inc.com}

\author{Jijia Lai}
\affiliation{%
  \institution{Alibaba Cloud Computing\\Alibaba Group}
  \city{Hangzhou, Zhejiang}
  \country{China}
}
\email{laijijia@gmail.com}

\author{Xiaoming Wang}
\orcid{0009-0001-1050-8364}
\affiliation{%
  \institution{Alibaba Cloud Computing\\Alibaba Group}
  \city{HangZhou, Zhejiang}
  \country{China}
}
\email{xiaoming.wxm@alibaba-inc.com}

\renewcommand{\shortauthors}{Mo et al.}

\begin{abstract}
Browser agents built on LLMs can act in web interfaces, yet most remain confined to a single chat surface (e.g., a sidebar). This mismatch with real browsing can increase context-switching and reduce user control. We introduce \textbf{IntentWeave}, a design space of ten spatial paradigms for embedding agentic assistance across a browser, organized as a progressive entry ladder from micro-interventions to dedicated workspaces. We implement IntentWeave as a browser-extension prototype on the Alibaba Cloud website and compare three entry strategies in a within-subjects study (N=16). Workspace-heavy strategies reduced completion time but lowered perceived control; micro-only strategies preserved control but were often insufficient; a mixed sidecar approach achieved the highest satisfaction. We conclude with guidance for escalating and retreating agent surfaces without disrupting user agency.

\end{abstract}




\maketitle

\section{Introduction}
Large Language Models are rapidly evolving from passive text assistants into agentic systems that can perceive, plan, and act in digital environments~\cite{barua2024exploring,gur2023real}. This evolution has motivated browser-based AI assistants that help users complete web tasks such as navigation, comparison, configuration, and troubleshooting. However, many current systems remain anchored in a single-surface paradigm, typically a sidebar~\cite{kuznetsov2022fuse} or floating chat window~\cite{mei2025interquest}, adjacent to page content~\cite{kronhardt2025proactive}.

Single-surface copilots mismatch real browsing behavior. Users routinely switch between pages, compare options, revisit earlier states, and shift between exploratory and exploitative modes~\cite{white2009exploratory}. Static chat interfaces also introduce context-switching friction: users must translate what they see into prompts and mentally track progress across pages, which can be cognitively demanding~\cite{subramonyam2024bridging}. Human-in-the-loop research suggests browser agents should lower effort while keeping users oriented and in control at every step~\cite{yun2025interaction}. In practice, this implies that \emph{how} an agent appears (and escalates) can be as important as what it can do.

Recent work explores embedded paradigms beyond chat, including page-level LLM overviews for sensemaking~\cite{liu2024selenite}, contextual overlays and in-situ interventions~\cite{kronhardt2025fallacycheck}, and workspace-style canvases for synthesis and planning~\cite{guo2026protosampling}. These approaches show promise but raise new design tensions: subtle agents may be missed, while proactive takeovers can disrupt flow and reduce perceived control~\cite{tang2025dark}. What is missing is a compact framework that helps designers decide \emph{which surface, when}, and how to transition between surfaces without harming trust or usability.

We introduce \textbf{IntentWeave}, a framework for \textbf{multi-surface agent entry choreography} in web browsing. IntentWeave defines a design space of ten spatial paradigms from in-context micro-interventions to dedicated workspaces, and a progressive entry ladder that supports escalation and retreat. We implement the framework in a browser-extension prototype for an \textbf{Alibaba Cloud} case study and evaluate three entry strategies in a within-subject user study.

Our contributions are fourfold. (1) a design space of ten spatial paradigms for agentic web interaction; (2) a progressive entry ladder for escalating/de-escalating assistance; (3) a prototype integrated into an Alibaba Cloud console case study; and (4) formative evidence on trade-offs between efficiency, perceived control, and satisfaction across entry strategies.

\begin{figure*}[ht]
    \centering
    \includegraphics[width=\linewidth]{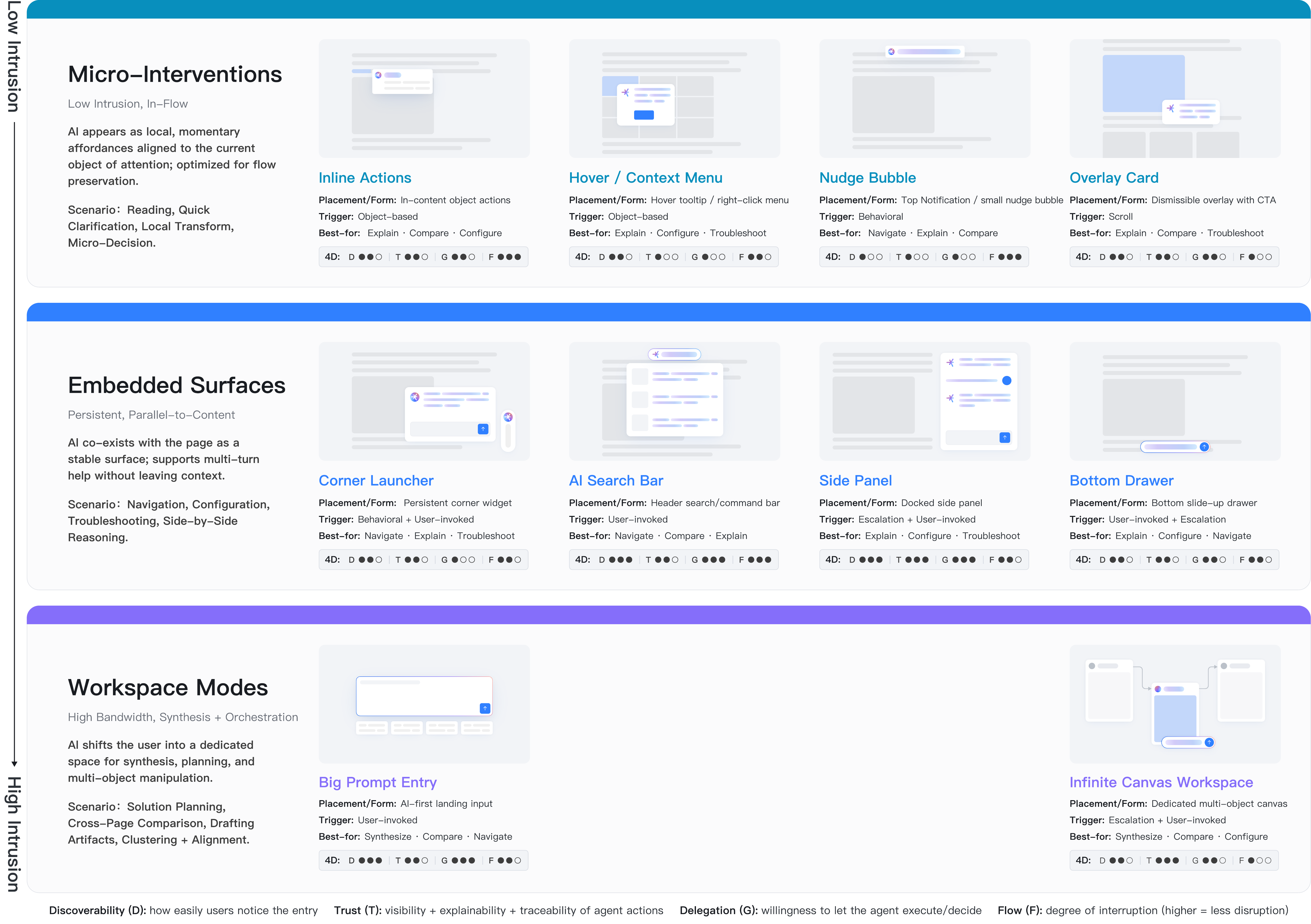}
    \caption{IntentWeave design space as a ten-paradigm \textbf{pro
    gressive entry ladder}}
    \label{fig:architecture}
\end{figure*}

\section{Design Space: Ten Spatial Paradigms for Agentic Web Interaction}
Where and how can an AI agent appear during a web session? To answer this, we surveyed recurring UI entry patterns across diverse web contexts (portal dashboards, documentation, and configuration consoles) and distilled them into \textbf{ten spatial paradigms}. These paradigms describe \emph{where} assistance is anchored and \emph{how} it captures attention, organizing them into a \textbf{progressive entry ladder} of three tiers: \textbf{Micro-interventions}, \textbf{Embedded Surfaces}, and \textbf{Workspace Modes} (Figure~\ref{fig:architecture}).

The goal is not to prescribe a single ``best'' UI, but to provide a vocabulary for \emph{entry choreography}. Each paradigm is characterized by its \textbf{placement}, \textbf{trigger} (e.g., object-based, behavioral, or user-invoked), and \textbf{best-for} intents. We compare them using four experience dimensions: \textbf{Discoverability (D)} of the entry point; \textbf{Trust/Legibility (T)} of the agent's actions; \textbf{Delegation (G)} potential; and \textbf{Flow (F)}, or the degree of interruption. This structure highlights a central tension: larger surfaces increase delegation but may reduce flow. Consequently, IntentWeave emphasizes \textbf{progressive entry}: the agent begins with the smallest effective surface and escalates only when user behavior (e.g., repeated backtracking) signals a need for greater support.

\subsection{Micro-interventions}
Micro-interventions are subtle, in-situ cues optimized for keeping users \emph{in flow}. In IntentWeave, micro-interventions are anchored to the current object of attention and are typically triggered by local context (object-based) or lightweight behavior (e.g., idle, repeated navigation). They are best for quick clarification, local transformations, and small decisions.

We include four micro-intervention paradigms. \textbf{Inline Actions} place small object-level affordances directly within content (e.g., \textit{Explain}, \textit{Compare}, \textit{Configure}) and are triggered by the presence of a relevant UI object. \textbf{Hover or Context Menus} provide just-in-time explanations through hover tooltips or right-click menus, making them well suited for unfamiliar parameters and troubleshooting scenarios. \textbf{Nudge Bubbles} use lightweight notifications, such as top banners, that are triggered by behavioral signals to improve discoverability without blocking ongoing interaction. Finally, \textbf{Overlay Cards} present dismissible call-to-action overlays, often triggered by scroll position or page progress, to suggest a concrete next step (e.g., ``Compare these services'').

Across these paradigms, users can quickly verify agent help in-place, which supports flow and incremental trust building. The main risk is \textbf{missed opportunity}: because micro-interventions are small, users may overlook them or find them insufficient for multi-step tasks (e.g., solution planning or cross-page comparison).

\begin{figure*}[t]
  \centering
  \includegraphics[width=\textwidth]{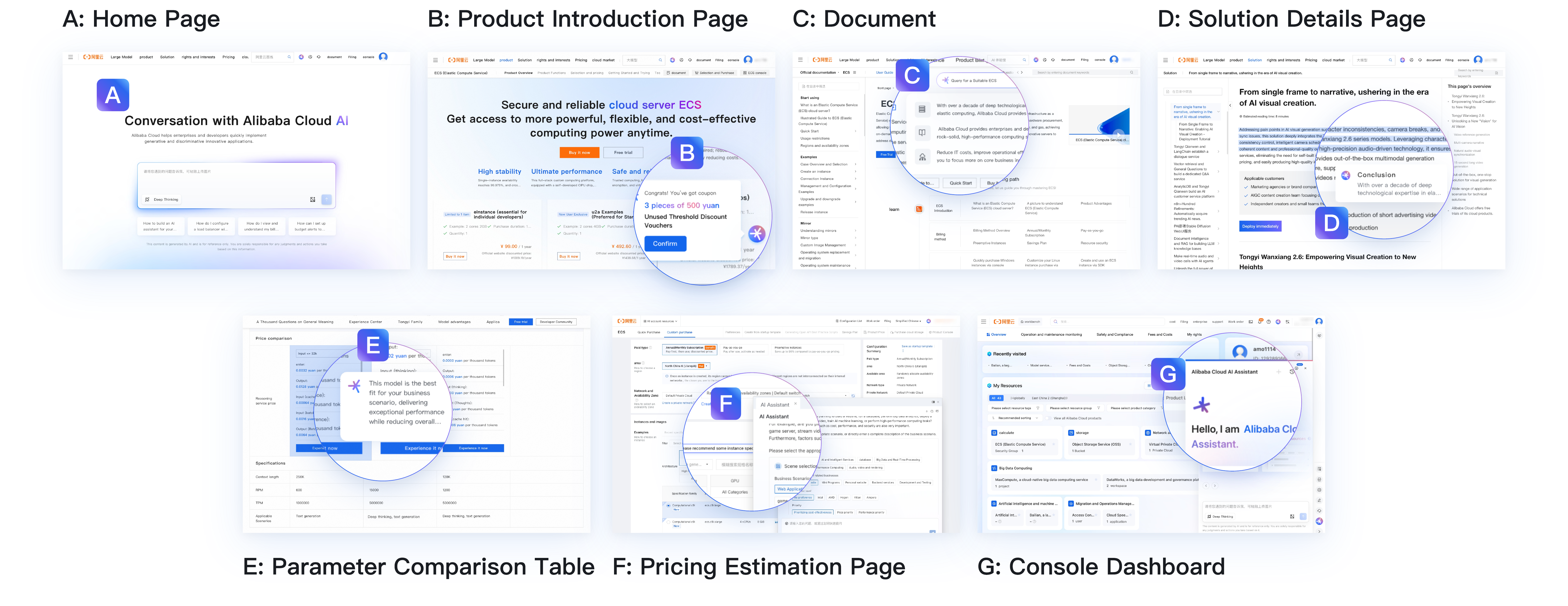}
  \caption[IntentWeave AI entry patterns]{IntentWeave on Alibaba Cloud website.
  \textbf{(A) Home Page:} Pre-initialized assistant dialogue; returning users are greeted with a dialog preloaded with inferred intent or shortcuts.
  \textbf{(B) Product Introduction Page:} Context-aware widget prompt; AI detects relevant offerings and displays coupons or promos based on product context.
  \textbf{(C) Document:} Search-driven guidance and intent prediction; AI enhances search by predicting task intent and suggesting relevant documentation.
  \textbf{(D) Solution Details Page:} Inline highlights and expandable summary; users can highlight text to receive AI-generated summaries and glossary explanations.
  \textbf{(E) Parameter Comparison Table:} Hover-triggered tooltip analysis; hovering over model names shows tailored pros/cons based on user profile and use case.
  \textbf{(F) Pricing Estimation Page:} Contextual overlay interaction; AI overlay allows conversational queries about pricing, discounts, or suggestions.
  \textbf{(G) Console Dashboard:} Floating assistant icon and side panel; users can query for diagnostics or resource status via a quick-access floating UI.}
  \label{fig:casestudy}
\end{figure*}

\subsection{Embedded Surfaces}
Embedded surfaces provide a persistent, parallel-to-content interface where the agent can hold state, show rationale, and support multi-turn guidance without fully taking over the page. They are best when tasks extend beyond a single field or page and the user benefits from continuous support (e.g., navigation across console pages, parameter configuration, troubleshooting).

IntentWeave includes four embedded-surface paradigms. \textbf{Corner Launcher} is a persistent entry widget that improves discoverability while remaining unobtrusive and remaining explicitly user-invoked. \textbf{AI Search Bar} embeds agent interaction into a familiar header search or command bar, supporting fast navigation and comparison-oriented queries. \textbf{Side Panel} is a docked surface that supports richer explanations, action previews, and an interaction trace, and commonly appears through escalation and/or direct user invocation. Finally, \textbf{Bottom Drawer} provides expandable assistance without permanently occupying lateral space and works well for short multi-step guidance and parameter inspection.

Compared to micro-interventions, embedded surfaces typically increase discoverability and legibility because users can see the agent's ongoing context and ``what it is doing.'' The trade-off is moderate disruption: these surfaces compete for attention and screen space, so they should be easily minimized and invoked contextually.

\subsection{Workspace Modes}
Workspace modes shift interaction into an AI-driven workspace for synthesis, planning, and multi-object manipulation. They are most appropriate when the user's intent expands beyond a single page (e.g., cross-page comparison, solution planning, drafting artifacts) or when tasks require aggregating scattered information. We include two workspace paradigms: \textbf{Big Prompt Entry} offers an AI-first landing input that encourages users to express high-level intent (synthesize/compare/navigate) before acting. \textbf{Infinite Canvas Workspace} provides a dedicated multi-object canvas for clustering, alignment, and assembling solution artifacts. It supports the highest delegation potential because it can host structured intermediate representations (e.g., comparison tables, multi-service architectures).

Workspace modes can improve efficiency for complex tasks, but they introduce the largest risk to flow: a workspace takeover can feel disorienting if invoked prematurely. Therefore, IntentWeave treats workspace entry as an \textbf{opt-in capstone}: transitions should be justified (``why this view now''), consented to, and reversible (clear return path to the originating page).

\begin{table*}[t!]
    \centering
    \resizebox{\textwidth}{!}{%
    \begin{tabular}{llp{7cm}l}
\toprule
\textbf{Task Name} & \textbf{Page/Context} & \textbf{Primary Goal} \\
\midrule
\textbf{T1: Solution Exploration} & Solution canvas page & Design a system architecture \\
\textbf{T2: Product Comparison} & Service info pages & Choose between two cloud services \\
\textbf{T3: Parameter Explanation} & Configuration form & Understand a specific setting\\
\textbf{T4: Troubleshooting \& Support} & Console/support page & Resolve an error or issue\\
\textbf{T5: Pricing Estimation} & Pricing calculator page & Calculate usage cost estimate\\
\bottomrule
    \end{tabular}
    }
    \caption{
    Representative tasks in the cloud portal prototype, with their context and user goal. 
    }
    \label{tab:scenario-ai-needs}
\end{table*}

\section{Case Study: IntentWeave on Alibaba Cloud}
We applied IntentWeave to the production Alibaba Cloud web portal shown in~\autoref{fig:casestudy} as our primary case study. We selected Alibaba Cloud because it represents the pinnacle of complex web-based system management, serving a massive user base ranging from students to enterprise architects. The platform's interface encompasses hundreds of products like Elastic Compute Service (ECS) and Object Storage Service (OSS), making it an ideal testbed for evaluating adaptive AI paradigms. Unlike simplified mockups, the Alibaba Cloud console requires users to navigate intricate dependencies (e.g., VPC configurations, security group rules) where a static \emph{one-size-fits-all} chatbot often fails to provide context-aware utility.

By injecting AI entry points directly into the live Alibaba Cloud DOM, we enabled multiple spatial paradigms that respond to the specific ``density'' of the user's task. For instance, while a simple inline hint suffices for explaining an ECS instance type, a complex architecture design requires an expanded Solution Canvas. This integration allows us to explore how AI can augment a real-world, commercial-grade interface without requiring the cloud provider to fundamentally redesign their legacy frontend. ~\autoref{tab:scenario-ai-needs} summarizes six representative scenarios used to construct tasks for evaluation.

\section{Evaluation and Results}
A within-subjects study was conducted in the organization to evaluate how the three entry strategies in ladder layer affect performance.

\subsection{User Study Specifications}

\subsubsection{Experimental Design}
We conducted a one-factor within-subjects experiment with 16 participants to evaluate the tasks detailed in~\autoref{tab:scenario-ai-needs}. The study compared three conditions corresponding to the IntentWeave ladder tiers: \textbf{(1) Micro-only}, where the agent provides only in-context hints without escalating; \textbf{(2) Mixed}, a ``companion'' approach utilizing side-panels alongside prompts; and \textbf{(3) Workspace-heavy}, where the agent aggressively escalates to full-page workspaces upon detecting complex intent.

\subsubsection{Procedure and Metrics}
To mitigate order effects, conditions were counterbalanced using a Latin square. Each participant performed 3 commands per task (15 total). Task-condition pairings were rotated across participants to prevent confounding task difficulty with assistance mode.

We recorded objective measures: \textbf{Task Completion Time} (s) and \textbf{Success Rate} (\%). Subjective experience was measured via post-task questionnaires on a 5-point Likert scale regarding \textbf{Overall Satisfaction}. Semi-structured interviews were conducted post-session to gather qualitative feedback on intrusiveness and trust.

\subsection{Key findings}
\begin{table*}[t]
\centering
\small
\begin{tabular}{llccc}
\toprule
\textbf{Task Context} & \textbf{Metric} & \textbf{Micro-only} & \textbf{Mixed} & \textbf{Workspace-heavy} \\
\midrule
\textbf{T1: Solution Exploration} & Completion Time (min) & 12.8 & 7.3 & 7.1 \\
                                  & Task Success (\%)   & 6.3  & 25.0 & 56.2 \\
                                  & Satisfaction        & 2.5   & 3.6   & 3.6 \\
\midrule
\textbf{T2: Product Comparison}   & Completion Time (min) & 3.5 & 3.6 & 3.1 \\
                                  & Task Success (\%)   & 87.5  & 100.0 & 100.0 \\
                                  & Satisfaction        & 3.6   & 4.6   & 4.4 \\
\midrule
\textbf{T3: Parameter Explanation}& Completion Time (min) & 1.8 & 2.4 & 3.8 \\
                                  & Task Success (\%)   & 100.0 & 100.0 & 87.5 \\
                                  & Satisfaction        & 4.8   & 4.2   & 3.0 \\
\midrule
\textbf{T4: Troubleshooting}      & Completion Time (min) & 8.6 & 4.9 & 5.0 \\
                                  & Task Success (\%)   & 31.5  & 75.0 & 75.0 \\
                                  & Satisfaction        & 3.1   & 4.3   & 4.1 \\
\midrule
\textbf{T5: Pricing Estimation}   & Completion Time (min) & / & 2.2 & 2.0 \\
                                  & Task Success (\%)   & /  & 100.0 & 100.0 \\
                                  & Satisfaction        & /   & 4.7   & 4.0 \\
\bottomrule
\end{tabular}
\caption{Objective and subjective metrics across N=16 participants. The results underscore the necessity of a progressive entry ladder: Micro-only strategies maximized efficiency for simple tasks (T3) but proved highly insufficient for complex workflows (T1, T4). Conversely, Workspace-heavy interventions reduced completion times and increased success rates for complex exploration, while the Mixed approach achieved the most consistent balance, yielding the highest overall satisfaction across contexts.}
\label{tab:results-expanded}
\end{table*}

As shown in~\autoref{tab:results-expanded}, we observed a consistent interaction between task complexity, efficiency, and satisfaction: \textbf{Efficiency:} Workspace-heavy strategies tended to reduce completion time by offloading multi-step work into dedicated workspaces (e.g., 7.1 min for T1), while Micro-only was faster for localized tasks but severely slowed down complex synthesis (12.8 min for T1). \textbf{Effectiveness:} Task success depended heavily on context. Failures typically reflected insufficient scaffolding for complex exploration or momentary disorientation from a large UI transition (Workspace-heavy). \textbf{Satisfaction:} Micro-only yielded the least disruption, but some participants described it as ``too subtle'' or ``not enough help''. Workspace-heavy improved speed for complex tasks but sometimes penalized satisfaction when the UI changed abruptly. Mixed consistently yielded the highest overall satisfaction: participants reported it felt like a \textit{sidekick} that was visible without taking over the workflow.

\subsection{Implications for design}
These findings support three practical implications for browser-agent UI:
(1) \textbf{Use micro-interventions as gateways} that can expand into an embedded surface when users need more depth.
(2) \textbf{Treat workspaces as opt-in power tools} for cross-page synthesis and planning, with explicit justification and a strong``return path.''
(3) \textbf{Design for reversible escalation:} visible controls to minimize/close, plus a lightweight activity trace, help preserve user orientation during proactive assistance.

\section{Limitations and Future Work}
This work is formative and has several limitations. First, the evaluation used a controlled prototype and short scripted tasks; real cloud-console work spans longer sessions, interruptions, and multi-day workflows. Second, our three conditions vary both spatial surface and escalation behavior; future studies could isolate surface effects from proactivity policy. Third, we focused on entry choreography rather than model failures; in production, mistakes may be tolerated differently depending on surface prominence (e.g., a wrong full-screen takeover may feel more costly than a wrong tooltip). Future work should test adaptive policies that personalize``boldness'' over time and validate IntentWeave in longitudinal field deployments.

\section{Conclusion}
IntentWeave reframes browser copilots as multi-surface agents whose effectiveness depends not only on model capability but also on \emph{entry choreography}. We contribute a design space of ten spatial paradigms and a progressive entry ladder for escalation and retreat, implemented in a cloud-portal prototype. A within-subject study suggests that micro-only approaches preserve control, workspace-heavy approaches improve efficiency, and mixed sidecar strategies best balance satisfaction and agency. We hope IntentWeave helps designers build browser agents that appear at the right time, in the right form, and know when to step back.

\bibliographystyle{ACM-Reference-Format}
\bibliography{sample-base}

\end{document}